\documentclass[12pt]{iopart}
\expandafter\let\csname equation*\endcsname\relax
\expandafter\let\csname endequation*\endcsname\relax
\usepackage{siunitx}
\usepackage[english]{babel}
\usepackage{amsfonts}
\usepackage[T1]{fontenc}
\usepackage[latin9]{inputenc}
\usepackage{geometry,mathtools}
\usepackage{color}
\usepackage{textcomp}
\usepackage{amssymb}
\usepackage{graphicx}
\usepackage[numbers]{natbib}
\usepackage[dvipsnames]{xcolor}

\usepackage{amsmath}

\begin{document}

\title{Heat current control in trapped BEC}

\author{C. Charalambous$^1$, M.A. Garcia-March$^1$,  M. Mehboudi$^1$, and M. Lewenstein$^{1,2}$}
\address{$^1$ICFO -- Institut de Ciencies Fotoniques, The Barcelona Institute of Science and Technology, Av. Carl Friedrich Gauss 3, 08860 Castelldefels (Barcelona), Spain}
\address{$^2$ICREA, Pg. Lluis Companys 23, 08010 Barcelona, Spain}

\begin{abstract}
We investigate the heat transport and the control of heat current among two spatially separated trapped Bose-Einstein Condensates
(BEC), each of them at a different temperature. To allow for heat transport among the two independent BECs we consider a link made of two harmonically trapped impurities, each of them interacting with one of the BECs. Since the impurities are spatially separated, we consider long-range interactions between them, namely a dipole-dipole coupling. 
We study this system under theoretically suitable and experimentally feasible assumptions/parameters. The dynamics of these impurities is treated within
the framework of the {\it quantum Brownian motion} model, where the excitation
modes of the BECs play the role of the {\it heat bath}. We address the dependence of heat current
and current-current correlations on the physical parameters of the
system. Interestingly, we show that heat rectification,
i.e., the unidirectional flow of heat, can occur in our system, when  a periodic driving on the trapping frequencies
of the impurities is considered. Therefore, our system is a possible
setup for the implementation of a phononic circuit. Motivated by recent
developments on the usage of BECs as platforms for quantum information
processing, our work offers an alternative possibility to use this
versatile setting for information transfer and processing, within
the context of phononics, and more generally in quantum thermodynamics. 
 \end{abstract}

\section{Introduction}
Control of heat transport has enormous potential applications,  beyond the traditional ones  in thermal insulation and efficient heat dissipation. 
It has been suggested as a resource for information processing, giving rise to striking technological developments. A series of smart heat current control devices, such as thermal
diodes~\cite{2002Terraneo,2004Li,2005Li,2006Hu}, thermal transistors
\cite{2006Li}, thermal pumps~\cite{2006Segal,2007Marathe,2010Ren}, thermal logic gates~\cite{2007Wang} or thermal memories~\cite{2008Wang-1},
have been proposed in the past decade. A key underlying idea in some of these devices is that heat transport associated  to phonons can realistically  be used to carry and process information. The science and engineering of 
heat manipulation and information processing using phonons is a brand new subject,
termed phononics~\cite{2008Wang,2012Li,2011Dubi}.
In the emerging field of quantum thermodynamics, the issues of heat transfer and heat
rectification are basic ingredients for the understanding and designing
heat engines or refrigerators at nanoscales. 
Besides the technological interests, highly controlled platforms for heat transport can potentially enable the study of fundamental theoretical questions, which will shed light on the study of thermodynamics of nonequilibrium systems \cite{2019Binder}.
%
In particular, heat conduction in low-dimensional systems has attracted
a growing interest  because of its multifaceted
fundamental importance in statistical physics, condensed matter physics,
material science, etc~\cite{2006Giazotto,2008Dhar-1,2003Lepri-1,2012Li-1}. 

In past years, advanced experimental techniques  have allowed for
the miniaturization of heat transport platforms down to the mesoscopic/microscopic
scale. Experimentally, a nanoscale solid state thermal rectifier using
deposited carbon nanotubes has been realized recently~\cite{2006Chang},
and a heat transistor---heat current of electrons controlled
by a voltage gate---has also been reported~\cite{2007Saira}. Furthermore,
it has been shown that thermal rectification can appear in a
two-level system asymmetrically coupled to phonon baths~\cite{2005Segal,2005Segal-1}.
In general, various models for thermal rectifiers/diodes that
allow heat to flow easily in one direction have been proposed~\cite{2002Terraneo,2004Li,2005Li,2006Hu,2005Segal,2006Lan,2005Hu,2007Yang}.
Among such platforms, interesting examples are those based on ultracold atomic systems, both at the theoretical
level~\cite{2018Wang,2016Jaramillo,2017Ye,2018Li} as well as at
the experimental one~\cite{2013Brantut}. In these works, the
paradigm of the model used is that of tailored time dependent protocols performing a heat cycle (often an Otto cycle). A cycle here consist of a series of steps performed in finite time where thermodynamic parameters are controlled such that heat is transferred from one bath into another. Frequently, the energy that
one needs to spend in implementing the aforementioned protocols is
too large for the amount of heat transferred. In addition, in such
a heat cycle, the Hamiltonian must be modified as adiabatically as
possible to avoid unwanted excitations. 
We remark that for the goals of the present work considering adiabatic changes in the Hamiltonian will require
long timescales, and therefore it can have drawbacks, such as exceeding the life
time of the BECs. Even though attempts have been made
to overcome this problem~\cite{2018Li,2013Torrontegui,2014Deffner},
our proposal circumvents it altogether by considering a rather different
paradigm. Instead of cycles, we consider  autonomous heat platforms, where the working medium
is permanently coupled to different baths and under the influence
of external time-dependent driving. 

\begin{figure}
	\includegraphics[width=0.99\columnwidth]{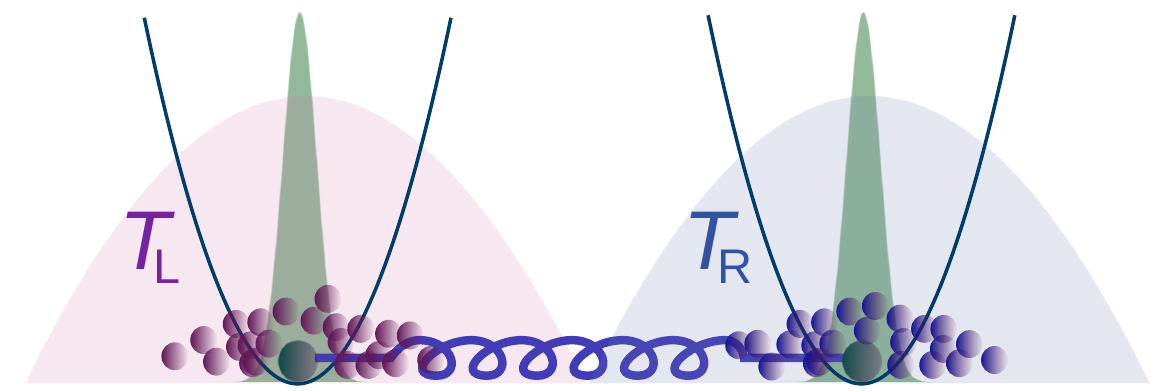} 
	\caption{Schematic of the system. The red and blue shaded profiles (and circles) represent the hot and cold Bose-Einstein condensates, respectively kept at $T_{\rm{L}}$ and $T_{\rm{R}}$. The green profiles (and circles) represent the two impurities trapped in their corresponding parabolic potentials, plotted as blue lines -- we omit the representation of the trap for the BECs. The spring-like blue line represents the long-range dipolar interaction among the impurities.  \label{Fig:fig1}}
\end{figure}

The primary aim of the present work is to design a method to transfer heat between two Bose-Einstein condensates (BECs)---with a comprehensive analytical description. The BECs are confined in independent one-dimensional parabolic traps, and kept at a certain distance such that the two BECs do not spatially overlap. Our second aim is to create thermal devices in our platform, specifically a heat rectifier.
In our platform, the working medium is constructed with two impurities, each of them immersed  in one of these BECs. The two impurities interact through long-range dipolar interactions (see Fig.~\ref{Fig:fig1} for an schematic of the system considered). Heat transfer through dipolar interactions has been also proposed recently in two parallel layers of dipolar ultracold Fermi gases~\cite{2018Renklioglu}. 
We first show that the Hamiltonian of the impurities in their corresponding
harmonically trapped BECs can be cast as that of two bilinearly
coupled Quantum Brownian particles interacting bilinearly with the Bogoliubov excitations of each BEC---which play the role of heat baths for the corresponding impurities. We analytically {\it derive} the spectral density (SD)
of the system and construct the quantum Langevin equations describing the out-of-equilibrium dynamics of the coupled impurities \cite{2019Lampo}. 
We emphasize that, following this procedure we avoid approximations involved in an alternative conventional approach based on Lindblad master equations, such as the Born-Markov approximation and the rotating wave approximation. 
We solve the quantum Langevin equations and find the covariance matrix of the impurities.
We henceforth focus on our two main aims. Firstly, we find exactly the steady state heat current between the BECs, and study how it can be manipulated by controlling relevant parameters, such as the trapping frequencies, dissipation strengths, and the physical distance of the BECs. Secondly, we introduce a periodic driving
on the trapping frequency of one impurity and show that our
setup can be used as a thermal rectifier. The setup that we present here can be used for implementing other thermal devices as well. 


The paper is structured as follows: In section \ref{sec:Ham}, we introduce the
Hamiltonian of our system, present the main assumptions involved,
and rewrite the Hamiltonian in a form analogous to that of two coupled brownian particles. In section  \ref{sec:QLE} we derive the Generalized Langevin
equations of motion for the two impurities and study
their solution in both static and periodically driven
scenarios. In section \ref{sec:mes} we present the relevant quantities of interest
and in section \ref{sec:results} we present our main results, both for heat transfer and rectification. Finally we summarize the results and conclude in section  \ref{sec:conc}. 
\section{The Model}
\label{sec:Ham}

We consider a system composed of two interacting impurities of the same mass $m$. Each one of these impurities is
embedded in a different BEC, which we label as L or R because they are trapped in parabolic potentials of frequencies  $\Omega_{\rm{B,L}}$ and $\Omega_{\rm{B,R}}$, respectively,  but one trapping potential is centered around a minima located in $\mathbf{d}_{\rm{L}}$ and the other around 
$\mathbf{d}_{\rm{R}}$, with certain distance among them.  Each BEC has, respectively,  $N_{\rm{L}}$
and $N_{\rm{R}}$ interacting atoms of the same species with mass $m_{\rm{B}}$. The two impurities are also trapped in a parabolic potential of frequencies $\Omega_{\rm{L}}$ and $\Omega_{\rm{R}}$ located around the same minima as their corresponding BEC.  
We allow for the scenario of these impurities being driven by an external
periodic force. The  two BECs do not overlap among  themselves (we will specify later the condition over the distance between the minima of the trapping potential).  With this, the Hamiltonian describing this setting is 
\begin{equation}
H_{\rm{Tot}}=H_{\rm{S}}+\sum_{j\in\left\{\rm{L,R}\right\} }\left(H_{\rm{B},j}+H_{\rm{BB},j}\right)+\sum_{j\in\left\{\rm{L,R}\right\} }H_{\rm{Int.},j}+H_{\rm{Drive}}\left(t\right)\label{eq: Initial Ham},
\end{equation}
where 
\begin{equation}
H_{\rm{S}}=\sum_{\begin{array}{c}
i,j\in\left\{\rm{L,R}\right\} ,\\
i\neq j
\end{array}}\frac{\mathbf{p}_{j}^{2}}{2m}+\mathbf{V}_{j}\left(\mathbf{x}_{j}\right)+H_{\rm{int.}},
\end{equation}
\begin{equation}
H_{\rm{B},j}=\int\mathbf{\Psi}_{j}^{\dagger}\left(\mathbf{x_{\rm{B}}}\right)\left[\frac{\mathbf{q}_{j}^{2}}{2m_{\rm{B}}}+\mathbf{V}_{\rm{B}}^{\left(j\right)}\left(\mathbf{x_{\rm{B}}}\right)\right]\mathbf{\Psi}_{j}\left(\mathbf{x_{\rm{B}}}\right)\rm{d}\mathbf{x_{\rm{B}}},
\end{equation}
\begin{equation}
H_{\rm{BB},j}=g_{\rm{B}}^{\left(j\right)}\int\mathbf{\Psi}_{j}^{\dagger}\left(\mathbf{x_{\rm{B}}}\right)\mathbf{\Psi}_{j}^{\dagger}\left(\mathbf{x_{\rm{B}}}\right)\mathbf{\Psi}_{j}\left(\mathbf{x_{\rm{B}}}\right)\mathbf{\Psi}_{j}\left(\mathbf{x_{\rm{B}}}\right)\rm{d}\mathbf{x_{\rm{B}}},
\end{equation}
\begin{equation}
\begin{array}{c}
H_{\rm{Int.},j}=g_{\rm{IB}}^{(j)}\int\mathbf{\Psi}_{j}^{\dagger}\left(\mathbf{x_{\rm{B}}}\right)\mathbf{\Psi}_{j}\left(\mathbf{x_{\rm{B}}}\right)\delta\left(\mathbf{x}_{j}-\mathbf{x_{\rm{B}}}\right)\rm{d}\mathbf{x_{\rm{B}}},\\
=g_{\rm{IB}}^{(j)}\mathbf{\Psi}_{j}^{\dagger}\left(\mathbf{x}_{j}\right)\mathbf{\Psi}_{j}\left(\mathbf{x}_{j}\right),
\end{array}
\end{equation}
where $\mathbf{x}$ and $\mathbf{x_{\rm{B}}}$ are the three-dimenional  position operators
of the impurity and the bosons respectively.  We assume contact interactions among the
bosons and between the impurity and the bosons, with strength given
by the coupling constants $g_{\rm{B}}^{\left(j\right)}$ and $g_{\rm{IB}}^{(j)}$
respectively. 
Here, $H_{\rm{int.}}$ denotes the interaction Hamiltonian
between the two impurities, which we will specify later. For the rest of the paper we assume that the trapping frequencies in two directions are much larger than in the third one. Therefore, the dynamics in those directions is effectively frozen and we can study the system in one dimension. From here on we assume that the minima of the potential are located at $d_{\rm{L}}=-d/2$ and $d_{\rm{R}}=d/2$.

We next  recast our
initial Hamiltonian (\ref{eq: Initial Ham}), in such a way as to
describe the motion of two interacting Quantum Brownian particles
in two separate BECs.  The procedure, which is the same as   that presented \cite{2018Lampoa} for the  case of a single impurity embedded in  a harmonically trapped BEC, can be summarized as follows: (i) we  make the BEC assumption,
i.e. that the condensate density greatly exceeds that of the above-condensate
particles, which results to only quadratic terms in the Hamiltonian;
 (ii)  we perform the Bogoliubov
transformation appropriate for the case of a harmonically trapped
BEC; (iii)  we solve the relevant Bogoliubov-de-Gennes equations,
in the limit of a one dimensional BEC that yields a Thomas-Fermi parabolic
density profile, as in~\cite{2004Petrov}; (iv) we finally assume
that the oscillations $x_{j}$ of each one of the impurities are much
smaller than the corresponding Thomas-Fermi radius $R_{j}$, 
\begin{equation}
x_{j}\ll R_{j},\label{eq:Linearity assumption}
\end{equation}
which physically implies that we study the dynamics of the impurities
in the middle of their corresponding bath traps, which allows us to
obtain a bilinear interaction of the position of the impurities and
the positions of their corresponding baths. 

These steps bring the Hamiltonian in the form
\begin{equation}
H_{\rm{Tot}}=H_{\rm{S}}\left(t\right)+\sum_{j\in\left\{\rm{L,R}\right\} }\widetilde{H}_{\rm{B},j}+\sum_{j\in\left\{\rm{L,R}\right\} }\widetilde{H}_{\rm{Int.},j},\label{eq: Final total Ham}
\end{equation}
where 
\begin{equation}
\widetilde{H}_{\rm{B},j}=\sum_{q}E_{q}^{\left(j\right)}b_{j,q}^{\dagger}b_{j,q},
\end{equation}
\begin{equation}
\widetilde{H}_{\rm{Int.},j}=\sum_{q}\hbar g_{q}^{\left(j\right)}x_{j}\left(b_{j,q}+b_{j,q}^{\dagger}\right),
\end{equation}
with 
\begin{equation}
E_{q}^{\left(j\right)}=\hbar\Omega_{\rm{B},j}\sqrt{q\left(q+1\right)},
\end{equation}
and 
\begin{equation}
g_{q}^{\left(j\right)}=\frac{g_{\rm{IB}}^{\left(j\right)}\mu_{j}}{\hbar\pi^{3/2}}\left[\frac{1+2q}{\hbar\Omega_{\rm{B},j}g_{\rm{B}}^{\left(j\right)}R_{j}^{3}}\right]^{1/2}\frac{\Gamma\left[\frac{1}{2}\left(1-q\right)\right]\Gamma\left[\frac{1}{2}\left(1+q\right)\right]\sin\left(\pi q\right)}{\left[q\left(q+1\right)\right]^{1/4}}.
\end{equation}
The chemical potential $\mu_{j}$ of the $j^{\rm{th}}$ bath is
\begin{equation}
\mu_{j}=\left(\frac{3}{4\sqrt{2}}g_{\rm{B}}^{\left(j\right)}N_{j}\Omega_{\rm{B},j}\sqrt{m_{\rm{B}}}\right)^{2/3},
\end{equation}
and the Thomas-Fermi radius $R_{j}$  of the $j^{\rm{th}}$ bath is
\begin{equation}
R_{j}=\sqrt{\frac{2\mu_{j}}{m_{\rm{B}}\left(\Omega_{\rm{B},j}\right)^{2}}}\,\,.
\end{equation}
At this point, it is important to note that Hamiltonian (\ref{eq: Final total Ham}),
which is indeed in the form of a Hamiltonian describing two coupled
quantum Brownian particles, was derived from the physical initial
Hamiltonian. Therefore,  this Hamiltonian lacks the 
renormalization term that guarantees that no ``runaway'' solutions
appear in the system, which is often included  in conventional open quantum system approaches~\cite{1962Coleman}. We are not going to introduce
such a term in order to guarantee the positivity of the Hamiltonian
and cast the system in the form of a minimal coupling theory with
$U\left(1\right)$ gauge symmetry.  Instead, we are  going to determine
in the next section a condition on the allowed parameters of the system
that will guarantee this. 

We assume dipole-dipole interactions among the two impurities.
From~\cite{2009Lahaye} we know that the form of dipole-dipole interaction
is the following:
\begin{equation}
H_{\rm{int.}}=A\frac{1}{r^{3}},\label{eq:Dipole-dipole}
\end{equation}
where $A=\frac{C_{dd}\left(1-3\left(\cos\theta\right)^{2}\right)}{4\pi}$, 
with 
\begin{equation}
C_{dd}=\mu^{2}\mu_{0},
\end{equation}
and $\mu_{0}=4\pi10^{-7}\frac{J}{A^{2}m}$ being the vacuum permeability,
while $\mu$ is the magnetic moment of the dipole. The angle $\theta$
is that formed between the axis of the two dipoles and it can
determined in experiments. Furthermore, in (\ref{eq:Dipole-dipole}),
$r=\left|\left(d_{\rm{R}}+x_{\rm{R}}\right)-\left(d_{\rm{L}}+x_{\rm{L}}\right)\right|$
is the distance between the two oscillators, with $d_{\rm{L}}$ and $d_{\rm{R}}$
being the centers of the trapping potentials of the two oscillators.
We then rewrite (\ref{eq:Dipole-dipole}) as 
\begin{equation}
H_{\rm{int.}}=Ad^{-3}\left|1+\frac{\left(x_{\rm{L}}-x_{\rm{R}}\right)}{d}\right|^{-3}\label{eq:Dipole - dipole 2}, 
\end{equation}
and we assume that the distance between the two oscillators centers
$d$ is much larger than the fluctuations $x_{\rm{L}},x_{\rm{R}}$
and hence the fluctuations difference $x_{\rm{L}}-x_{\rm{R}}$, i.e. $x_{\rm{L}},x_{\rm{R}}\ll d$.
Importantly, let us emphasize that this is not an additional assumption, because we assumed right
from the beginning that the two baths should not overlap. This indeed means that the sum of the Thomas-Fermi radius have to be smaller than
the distance between the impurities, $R_{\rm{L}}+R_{\rm{R}}<d$.
With the additional assumption we made before, namely that the impurities
oscillations are much smaller than their corresponding Thomas-Fermi
radius, Eq.~(\ref{eq:Linearity assumption}), then one concludes that $x_{\rm{L}},x_{\rm{R}}\ll d$.
One could also tackle the problem of interacting baths, by making
use of the work in~\cite{2012Li} where one needs
to consider the surface Green functions, and in this case then the
assumption $x_{\rm{L}},x_{\rm{R}}\ll d$ should be made
explicitly. 

Finally,  after expanding the binomial
series, Eq.~(\ref{eq:Dipole - dipole 2}) is rewritten as
\begin{equation}
H_{\rm{int.}}=Ad^{-3}\left(1-3\frac{\left(x_{\rm{L}}-x_{\rm{R}}\right)}{d}+6\frac{\left(x_{\rm{L}}-x_{\rm{R}}\right)^{2}}{d^{2}}\right).
\end{equation}
One can show that the first two terms in the parenthesis will not
contribute to the dynamics of the impurities, since they are linear
in the displacement operators $x_{\rm{L}},x_{\rm{R}}$ and hence they will appear
only as constants in the equations of motion that we will study later
on, which will be obtained from the Heisenberg equations of motion.
The third term then is expanded as
\begin{equation}
\widehat{H}_{\rm{int.}}=6A d^{-5}\left(x_{\rm{L}}^{2}-2x_{\rm{L}}x_{\rm{R}}+x_{\rm{R}}^{2}\right),
\end{equation}
and we absorb the terms with $x_{\rm{L}}^{2}$ and $x_{\rm{R}}^{2}$ in the
non-interacting part of the Hamiltonian by redefining the frequencies
as 
\begin{equation}
H_{\rm{S}}^{\rm{non-inter.}}=\frac{p_{\rm{L}}^{2}}{2m}+\frac{p_{\rm{R}}^{2}}{2m}+\frac{m}{2}\hat{\Omega}_{\rm{L}}^{2}x_{\rm{L}}^{2}+\frac{m},{2}\hat{\Omega}_{\rm{R}}^{2}x_{\rm{R}}^{2}
\end{equation}
where 
\begin{equation}
\begin{array}{c}
\hat{\Omega}_{\rm{L}}^{2}=\Omega_{\rm{L}}^{2}+6A d^{-5},\\
\hat{\Omega}_{\rm{R}}^{2}=\Omega_{\rm{R}}^{2}+6A d^{-5}.
\end{array}
\end{equation}
From here on we omit the tilde in the frequencies to avoid unnecessary complications in the nomenclature.  Therefore we can rewrite the interaction as
\begin{equation}
\widetilde{H}_{\rm{int.}}=\kappa x_{1}x_{2},
\end{equation}
which models a spring-like interaction among the two impurities with  
\begin{equation}
\kappa=\frac{12C_{dd}\left(1-3\left(\cos\theta\right)^{2}\right)}{4\pi d ^{5}}.
\end{equation}
The angle $\theta$ can be experimentally controlled as in~\cite{2018Ilzhofer}. 
However, it is important to note here that the results that we present
in this paper are valid even if the angle between the dipoles can
not be experimentally controlled, but rather an average over the angle
is considered. The constants in the interacting Hamiltonian as well
as the power dependence on the relative distance will be different,
but in the limit we consider, i.e. when $x_{\rm{L}},x_{\rm{R}}\ll d$,
qualitativley the results will be the same, with the difference being
that the distance at which the oscillators should be kept will change
\cite{2016Reifenberger}. We also note that, in most ultracold dipolar gases, the dipolar interaction is present
together with the short range interactions arising from low angular
momentum scattering~\cite{2009Lahaye,2008Giorgini}. Usually the
latter is dominant and a Feshbach resonance is needed to probe the
regimes where dipolar effects are prominent. However in the setting
that we will consider in this work, as we maintain the two BEC baths
spatially separated, the effect of the short range interactions is
negligible, such that dipole interaction is the main process through
which current is transferred. 

Finally, there are a number of ways to drive our system, either by
driving degrees of freedom of the central system, or degrees of freedom
of the environment, or their coupling. In this work we focus on the
first case. There are basically two types of driving that one could
consider and would maintain the quadratic form of the Hamiltonian
such that an analytic solution to the resulting equations of motion
can be obtained. First we could consider applying a periodically driven
ramp potential on the central particles degrees of freedom only, of
the form $H_{\rm{Drive}}\left(t\right)=\Theta\left(t-t_{0}\right)\mathbf{f}^{T}\left(t\right)\mathbf{X}\left(t\right)$
where $\mathbf{X}\left(t\right)=\left(
x_{1}\left(t\right), 
x_{1}\left(t\right)
\right)^\intercal$ and $\mathbf{f}\left(t\right)$ is some periodic function, e.g. $\mathbf{f}\left(t\right)=\mathbf{f}_{0}e^{-i\omega_dt}+c.c.$
with $t_{0}$ the time at which the driving begins, $\Omega_d$
the driving frequency and $\mathbf{f}_{0}$ a complex valued constant
column vector, $\mathbf{f}_{0}=\left(
f_{0,\rm{L}},
f_{0,\rm{R}}
\right)^\intercal$. This type of driving was considered in~\cite{2011Agarwalla}, and
can represent the force exerted on the system by a time dependent
electromagnetic field. For this kind of driving however, it was shown in~\cite{2007Marathe} that with the setup we assume above, neither
a heat engine nor a heat pump can be constructed. Furthermore, our
goal is to construct a phononic diode with our setup which means that
the driving should be able to induce a unidirectional flow of heat
current, and this is not the case for this type of driving. Therefore,
in our work we consider the only other possible type of driving on
our system that maintains our Hamiltonian in a quadratic form, i.e. 
\begin{equation}
H_{\rm{Drive}}\left(t\right)=\Theta\left(t-t_{0}\right)\frac{1}{2}\mathbf{X}^{T}\cdot\mathbf{V}\left(t\right)\mathbf{X},\label{eq:Driven Ham-1}
\end{equation}
where the driving is either on the trapping frequency of the central
oscillators or on their in-between coupling. It was recently shown in~\cite{2019Riera-Campeny} that in such escenario one can observe
the appearance of phenomenon of heat rectification, and also there
is the potential to construct a heat engine as it was shown in~\cite{2018Hofer},
by introducing a coherently driven coupling between the two oscillators.
We also assume the driving to be periodic, $\mathbf{V}\left(t+\tau\right)=\mathbf{V}\left(t\right)$ 
with $\tau$ being the time period, such that it can be Fourier expanded
as 
\begin{equation}
\mathbf{V}\left(t\right)=\sum_{k}\mathbf{V}_{k}e^{ik\Omega_dt},
\end{equation}
where $\Omega_d=1/\tau$ being the driving frequency. This type
of coupling could also be implemented by a laser.

\section{Quantum Langevin equations}
\label{sec:QLE}

%


Let us now derive  the equations of motion for the two impurities.  First we write the Heisenberg equations of motion for the
bath 
\begin{eqnarray}
\frac{db_{k,j}\left(t\right)}{dt}&=\frac{i}{\hbar}\left[H_{\rm{Tot}},b_{k,j}^{\dagger}\left(t\right)\right]=-i\omega_{k,j}b_{k,j}\left(t\right)-g_{k,j}x_{j}\left(t\right),\nonumber\\
\frac{db_{k,j}^{\dagger}\left(t\right)}{dt}&=\frac{i}{\hbar}\left[H_{\rm{Tot}},b_{k,j}\left(t\right)\right]=i\omega_{k,j}b_{k,j}^{\dagger}\left(t\right)+g_{k,j}x_{j}\left(t\right),
\end{eqnarray}
and impurity particles
\begin{eqnarray}
\frac{dx_{j}\left(t\right)}{dt}\hspace{-0cm}=\frac{i}{\hbar}\left[H_{\rm{Tot}},x_{j}\left(t\right)\right]=\frac{p_{j}\left(t\right)}{m},\label{eq:impurities position eom}
\\
%
\frac{dp_{j}}{dt}\left(t\right)\hspace{-0cm}=\frac{i}{\hbar}\left[H_{\rm{Tot}},p_{j}\left(t\right)\right]\label{eq:impurities momentum eom}
\\
\hspace{1.3cm}=-m(\Omega_{j}^{2}+\sum_{k}V_{k,jj}e^{ik\omega_dt})x_{j}\left(t\right)
\\
\hspace{1.3cm}-(\kappa+\sum_{k}V_{k,jj}e^{ik\omega_dt})x_{q}\left(t\right)-\hbar\sum_{k}g_{k,j}\left(b_{k,j}\left(t\right)+b_{k,j}^{\dagger}\left(t\right)\right),\nonumber
\end{eqnarray}
where $j,q\in\left\{\rm{L,R}\right\} $ and $j\neq q$. We first solve
the bath particles equations of motion
\begin{eqnarray}
b_{kj}\left(t\right)&=b_{kj}\left(0\right)e^{-i\omega_{k}t}+\int_{0}^{t}\frac{g_{k,j}}{2}e^{i\omega_{k}\left(t-s\right)}x_{j}\left(s\right)\rm{d}s,\\
b_{kj}^{\dagger}\left(t\right)&=b_{kj}\left(0\right)e^{i\omega_{k}t}+\int_{0}^{t}\frac{g_{k,j}}{2}e^{-i\omega_{k}\left(t-s\right)}x_{j}\left(s\right)\rm{d}s,
\end{eqnarray}
and we replace these in the impurities equations of motion (\ref{eq:impurities position eom})--(\ref{eq:impurities momentum eom}),
to obtain
\begin{equation}
\hspace{-0cm}\ddot{x}_{j}+(\Omega_{j}^{2}+\sum_{k}V_{k,jj}e^{ik\omega_dt})x_{j}+(\kappa+\sum_{k}V_{k,jq}e^{ik\omega_dt})x_{q}-\int_{0}^{t}\lambda_{j}\left(t-s\right)x_{j}\left(s\right)\rm{d}s=\frac{1}{m}B_{j}\left(t\right),\label{eq:Generalized Langevin eq}
\end{equation}
where $B_{j}\left(t\right)$ plays the role of the stochastic force
and reads as
\begin{equation}
B_{j}\left(t\right)=\sum_{k}\hbar g_{k,j}\left(b_{kj}^{\dagger}e^{-i\omega_{k}t}+b_{kj}e^{i\omega_{k}t}\right).
\end{equation}
Here $\lambda_{j}\left(t\right)$ is called the susceptibility or noise
kernel.  In this setting it can also be identified as the self-energy
contributions coming from the bath, and it reads as
\begin{equation}
\lambda_{j}\left(t\right)=\frac{1}{m}\sum_{k}\hbar g_{k,j}^{2}\sin\left(\omega_{k}t\right)=\frac{1}{m}\int_{0}^{\infty}J_{j}\left(\omega\right)\sin\left(\omega t\right)\rm{d}\omega,
\end{equation}
with
\begin{equation}
J_{j}\left(\omega\right)=\sum_{k}\hbar g_{k,j}^{2}\delta\left(\omega-\omega_{k}\right),
\end{equation}
being the spectral density. Equation (\ref{eq:Generalized Langevin eq})
has the form of a Generalized Langevin equation which describes the
evolution of a system with memory and under the influence of a stochastic
force. These  terms, $B_{j}\left(t\right)$ and $\gamma_{j}\left(t\right)$,
contain all the relevant information about the baths. Furthermore,
let us assume the Feynman-Vermon initial state assumption,
i.e. that the initial conditions of the impurities and the bath oscillators
are uncorrelated,
\begin{equation}
\rho\left(0\right)=\rho_{\rm{S}}\left(0\right)\otimes\rho_{\rm{B}},\label{eq:static bath assumption}
\end{equation}
where $\rho\left(0\right)$ is the total density state, $\rho_{\rm{S}}\left(0\right)$
is the initial density state of the system and $\rho_{\rm{B}}$ is the
density state of the bath which is assumed to be thermal and hence
is a Gibbs state. Then, it can be shown that  the Fourier transform of $B_{j}\left(t\right)$
obeys the fluctuation dissipation relation 
\begin{equation}
\frac{1}{2}\left\langle \left\{ B_{j}\left(\omega\right),B_{q}\left(\omega'\right)\right\} \right\rangle =\delta_{jq}Im\left[\lambda_{j}\left(\omega\right)\right]\coth\left(\frac{\hbar\omega}{2k_{\rm{B}}T}\right)\delta\left(\omega-\omega'\right),\label{eq:Fluctuation Dissipation}
\end{equation}
where $\lambda_{j}\left(\omega\right)$ is the Fourier transform of
$\lambda_{j}\left(t\right)$ and obeys~\cite{2013Valido}
\begin{equation}
Im\left[\lambda_{j}\left(\omega\right)\right]=-\hbar\left(\Theta\left(\omega\right)-\Theta\left(-\omega\right)\right)J_{j}\left(\omega\right).
\end{equation}
Hence one concludes that, upon determination of  the spectral density $J_{j}\left(\omega\right)$, one
determines the influence of the baths on the impurities. In~\cite{2018Lampoa},
it was shown that in the continuous frequency limit, the spectral
density takes the following form
\begin{equation}
J_{j}\left(\omega\right)=m\tau_{j}\frac{\omega^{4}}{\Lambda_{j}^{3}}\Theta\left(\omega-\Lambda_{j}\right),
\end{equation}
where 
\begin{equation}
\tau_{j}=\frac{2g_{\rm{B}}^{\left(j\right)}}{m\widehat{\Omega}_{\rm{B},j}R_{j}^{3}}\left(\frac{\eta_{j}\mu_{j}}{\hbar\widehat{\Omega}_{\rm{B},j}}\right)^{2},\;\eta_{j}=\frac{g_{\rm{IB}}^{\left(j\right)}}{g_{\rm{B}}^{\left(j\right)}},\;\Lambda_{j}=\widehat{\Omega}_{\rm{B},j}.
\end{equation}
Note that the ultraviolent cutoff, that is usually
introduced to regularize the spectrum in the conventional QBM model,
is now given in terms of a physical quantity, the trapping
frequency of the potential well of the $j^{th}$ bath. Here  $\frac{\tau_{j}}{\Lambda_{j}^{3}}$
plays the role of a relaxation time. Heat transport
with a superohmic spectral density was considered for example in the
energy transport in the phenomenon of photosynthesis~\cite{2017Qin}.
Note that Eq. (\ref{eq:Generalized Langevin eq}) can be rewritten
in terms of the damping kernel $\gamma_{j}\left(t\right)$ which is
related to the susceptibility by $\frac{1}{m}\lambda_{j}\left(t\right)=-\frac{\partial}{\partial t}\gamma_{j}\left(t\right)$
\cite{2015Dhar}, as 
\begin{align}
\ddot{x}_{j}+\tilde{\Omega}_j^2x_{j}-\gamma_{j}\left(0\right)x_{j}+\tilde{\kappa}_q x_{q}+\frac{\partial}{\partial t}\int_{0}^{t}\!\!\gamma_{j}\left(t-s\right)x_{j}\left(s\right)\rm{d}s=\frac{1}{m}B_{j}\left(t\right),
\end{align}
with $\tilde{\Omega}_j^2=\Omega_{j}^{2}+\sum_{k}V_{k,jj}e^{ik\omega_dt}$ and $\tilde{\kappa}_q=\kappa+\sum_{k}V_{k,jq}e^{ik\omega_dt}$, 
where the Leibniz rule was used as in ~\cite{2019Charalambous}. Moreover,
in~\cite{2018Lampoa}, it was also shown that the form of the damping
kernel for such a spectral density reads as
\begin{align}
\gamma_{j}\left(t\right)&=\\
&\frac{\tau_{j}\left(6+3\left(\left(\widehat{\Omega}_{\rm{B},j}\right)^{2}t^{2}-2\right)\cos\left(\widehat{\Omega}_{\rm{B},j}t\right)\right)}{t^{4}\left(\widehat{\Omega}_{\rm{B},j}\right)^{3}}
+\frac{\tau_{j}\widehat{\Omega}_{\rm{B},j}\left(\left(\widehat{\Omega}_{\rm{B},j}\right)^{2}t^{2}-6\right)\sin\left(\widehat{\Omega}_{\rm{B},j}t\right)}{t^{3}\left(\widehat{\Omega}_{\rm{B},j}\right)^{3}}.\nonumber
\end{align}
In the limit  $t\rightarrow0$ this damping kernel becomes
\begin{equation}
\gamma_{j}\left(0\right)=\lim_{t\rightarrow0}\gamma_{j}\left(t\right)=\frac{\widehat{\Omega}_{\rm{B},j}\tau_{j}}{4}.
\end{equation}
In vector form the two coupled equations in (\ref{eq:Generalized Langevin eq})
read as
\begin{equation}
\ddot{\mathbf{X}}\left(t\right)+\mathbf{K}\cdot\mathbf{X}\left(t\right)+\frac{\partial}{\partial t}\int_{0}^{t}\mathbf{D}\left(t-s\right)\cdot\mathbf{X}\left(s\right)\rm{d}s=\frac{1}{m}\mathbf{B}^{T}\left(t\right)\mathbb{I},\label{eq:Vector form of eom}
\end{equation}
where $\mathbb{I}$ is the identity matrix and 
\begin{eqnarray*}
\hspace{-1cm}&\mathbf{X}\left(t\right)=\left(\begin{array}{c}
x_{\rm{L}}\left(t\right)\\
x_{\rm{R}}\left(t\right)
\end{array}\right),\\
\hspace{-1cm}&\mathbf{K}=\left(\begin{array}{cc}
\Omega_{\rm{L}}^{2}+\sum_{k}V_{k,LL}e^{ik\Omega_dt}-\gamma_{\rm{L}}\left(0\right) & \kappa+\sum_{k}V_{k,LR}e^{ik\Omega_dt}\\
\kappa+\sum_{k}V_{k,RL}e^{ik\Omega_dt} & \Omega_{\rm{R}}^{2}+\sum_{k}V_{k,RR}e^{ik\Omega_dt}-\gamma_{\rm{R}}\left(0\right)
\end{array}\right),\\
\hspace{-1cm}&\mathbf{D}\left(t\right)=\left(\begin{array}{cc}
\gamma_{\rm{L}}\left(t\right) & 0\\
0 & \gamma_{\rm{R}}\left(t\right)
\end{array}\right),\,\,\mbox{and}\,\,\,\,\,
\mathbf{B}\left(t\right)=\left(\begin{array}{c}
B_{\rm{L}}\left(t\right)\\
B_{\rm{R}}\left(t\right)
\end{array}\right).
\end{eqnarray*}

\paragraph{Static case.}

From the system of coupled
equations (\ref{eq:Generalized Langevin eq}), one can now  identify the condition that guarantees the positivity
of the Hamiltonian in the static case, where 
$H_{\rm{Drive}}\left(t\right)=0$ . To this end one  first diagonalizes  the Hamiltonian, and then requires  that the normal
mode frequencies are positive. We diagonalize the Hamiltonian by making
the transformation $\mathbf{Q}=\mathbf{O}\cdot\mathbf{X}$, that brings
(\ref{eq:Vector form of eom}) into
\begin{equation}
\ddot{\mathbf{Q}}\left(t\right)+\mathbf{K}_{D}\cdot\mathbf{Q}\left(t\right)+\frac{\partial}{\partial t}\int_{0}^{t}\mathbf{D}_{D}\left(t-s\right)\cdot\mathbf{Q}\left(s\right)\rm{d}s=\frac{1}{m}\mathbf{B}_{D}^{T}\left(t\right)\underline{\underline{I}},
\end{equation}
where $\mathbf{D}_{D}\left(t\right)=\mathbf{O}\cdot\mathbf{D}\left(t\right)\cdot\mathbf{O}^{T}$,
$\mathbf{B}_{D}^{T}\left(t\right)=\mathbf{O}\cdot\mathbf{B}^{T}\left(t\right)$,
and the frequency matrix is diagonalized as 
\begin{equation}
\mathbf{K}_{D}=\mathbf{O}\cdot\mathbf{K}\cdot\mathbf{O}^{T}=\left(\begin{array}{cc}
\Omega_{\rm{L}}^{D} & 0\\
0 & \Omega_{\rm{R}}^{D}
\end{array}\right).
\end{equation}
The positivity of the Hamiltonian condition is then guaranteed by
requiring that $\{\Omega_{\rm{L}}^{D},\Omega_{\rm{R}}^{D}\}>0$ which in terms of
the original frequencies reads as 
\begin{eqnarray}
&\frac{1}{2}\Bigg[\gamma_{\rm{L}}\left(0\right)+\gamma_{\rm{R}}\left(0\right)-\Omega_{\rm{L}}^{2}-\Omega_{\rm{R}}^{2}+\!\Big[\!\left(\gamma_{\rm{L}}\left(0\right)+\gamma_{\rm{R}}\left(0\right)-\Omega_{\rm{L}}^{2}-\Omega_{\rm{R}}^{2}\right)^{2}
\label{eq:positivity condition}\\
&-4\left(\gamma_{\rm{L}}\left(0\right)\gamma_{\rm{R}}\left(0\right)-\kappa^{2}-\gamma_{\rm{L}}\left(0\right)\Omega_{\rm{R}}^{2}-\gamma_{\rm{R}}\left(0\right)\Omega_{\rm{L}}^{2}+\Omega_{\rm{R}}^{2}\Omega_{\rm{L}}^{2}\right)\Big]^{1/2}\Bigg]\!<\!0\nonumber.
\end{eqnarray}
This condition  guarantees that we do not have
negative renormalized normal frequencies in the system and hence
stability of the solution is guaranteed in the long-time limit. With this satisfied, we
are in a position to safely neglect the effects of $\gamma_{\rm{L}}\left(0\right)$
and $\gamma_{\rm{R}}\left(0\right)$ in the dynamics of the system. 

Upon rewriting the coupled equations of motion for the static case in a vector
form as above, and considering it expressed in terms of the susceptibilities
$\lambda_{\rm{L}}\left(t\right),\lambda_{\rm{R}}\left(t\right)$, one can obtain
its solution  by taking  the Fourier transform of
both sides
\begin{equation}
\mathbf{X}\left(\omega\right)=\mathbf{G}\left(\omega\right)\frac{\mathbf{B}\left(\omega\right)}{m},
\end{equation}
where
\begin{equation}
\mathbf{G}\left(\omega\right)=\left(-\omega^{2}\mathbf{I}+\mathbf{K}-\mathbf{L}\left(\omega\right)\right)^{-1},
\end{equation}
which is understood to play the role of a phonon Green function, and
$\mathbf{L}\left(\omega\right)$ is the Fourier transform of $\mathbf{L}\left(t\right)=\left(\begin{array}{cc}
\lambda_{\rm{L}}\left(t\right) & 0\\
0 & \lambda_{\rm{R}}\left(t\right)
\end{array}\right)$, with diagonal elements 
\begin{eqnarray}
&\hspace{2.5cm}\lambda_{j}\left(\omega\right)=Re[\lambda_{j}\left(\omega\right)]+iIm[\lambda_{j}\left(\omega\right)]\\
&\hspace{-2.2cm}=\frac{\tau_{j}\left(\Lambda_{j}^{4}+2\Lambda_{j}^{2}\omega^{2}+2\omega^{4}\left(i\pi+\log\left(-1+\frac{\Lambda_{j}^{2}}{\omega^{2}}\right)\right)\right)}{\pi^{2}}-i(\hbar\left(\Theta\left(\omega\right)-\Theta\left(-\omega\right)\right)J_{j}\left(\omega\right)),\nonumber
\end{eqnarray}
where the real part of the susceptibility was obtained through the
Kramers-Kronig relation $Re[\lambda_{j}\left(\omega'\right)]=\mathcal{H}[Im[\lambda_{j}\left(\omega\right)]]\left(\omega'\right)=\frac{1}{\pi}P\int_{-\infty}^{\infty}\frac{Im[\lambda_{j}\left(\omega\right)]}{\omega-\omega'}\rm{d}\omega$. Here $\mathcal{H}[\cdot]\left(\omega'\right)$ denotes the Hilbert
transform and $P$ the principal value. In
general, for the parameters we consider in our results,
it will always hold that $\Omega_{j}^{2}\gg Re[\lambda_{j}\left(\omega\right)]$
such that we safely neglect the effect of $Re[\lambda_{\rm{L}}\left(\omega\right)]$
and $Re[\lambda_{\rm{R}}\left(\omega\right)]$.

In terms of this solution of the static equations of motion, the positivity condition
(\ref{eq:positivity condition}), can  be interpreted in a different way: it  guarantees
that the phonon propagator, i.e. the Green function $\mathbf{G}\left(\omega\right)$,
has no poles in the lower half plane of the complex plane. This implies
that there are no divergencies in the integrals that will be performed
later on and will involve these Green functions~\cite{1980Rzazewski}. 

\paragraph{Driven case.}

Now we consider the case where a driving is applied on
the central system. In particular we assume that the driving is either
on the oscillators' frequencies or on their inbetween coupling, as
in \cite{2019Riera-Campeny}. The analytic treatment of this case is slightly more involved than the static one. We are now dealing with a periodic differential
equation, and the analysis of the stability of the long-time steady
state solution of the equations of motion is not straightforward. To be able to
perform the stability analysis, what one usually does is to convert
the periodic differential equation to a linear one by resorting to
Floquet theory. This is done by converting first all the terms in
the equation of motion into periodic ones, and then study the stability
of the Floquet-Fourier components of the resulting Green function.
The basic assumption that enables us to employ the Floquet formalism
is that even though some function $f\left(t\right)$ might not be
periodic, another function defined based on this one as $\mathbf{W}\left(t,\omega\right)=\int_{\mathbb{R}}dt'f\left(t-t'\right)e^{i\omega\left(t-t'\right)}$
will indeed be periodic. Furthermore, such periodic function can always
be expressed in terms of its Fourier components as $\mathbf{W}\left(t,\omega\right)=\sum_{k}\mathbf{B}_{k}\left(\omega\right)e^{ik\Omega_dt}$.
By performing these two transformations on the equations of motion, i.e. $\int_{\mathbb{R}}dt'e^{i\omega\left(t-t'\right)}$ 
and Fourier expanding, one obtains a set of equations for the Fourier
coefficients $\mathbf{A}_{k}\left(\omega\right)$ of $\mathbf{P}\left(t,\omega\right)=\int_{\mathbb{R}}dt'\mathbf{G}\left(t-t'\right)e^{i\omega\left(t-t'\right)}=\sum_{k}\mathbf{A}_{k}\left(\omega\right)e^{ik\Omega_dt}$
that can be selfconsistently solved. By following this procedure,
in \cite{2019Riera-Campeny} 
the authors were able to obtain expressions for these
coefficients 
\begin{eqnarray}
\hspace{-1.25cm}&\mathbf{A}_{0}\left(\omega\right)=\mathbf{G}\left(\omega\right)+\sum_{j\neq0}\mathbf{G}\left(\omega\right)\cdot\mathbf{V}_{j}\cdot\mathbf{G}\left(\omega+j\Omega_d\right)\cdot\mathbf{V}_{-j}\cdot\mathbf{G}\left(\omega\right)+\mathcal{O}\left(\mathbf{V}_{j}^{4}\right),\\
\hspace{-1.25cm}&\mathbf{A}_{k}\left(\omega\right)=-\mathbf{G}\left(\omega-k\Omega_d\right)\cdot\mathbf{V}_{k}\cdot\mathbf{G}\left(\omega\right)+\mathcal{O}\left(\mathbf{V}_{j}^{3}\right),
\end{eqnarray}
where $\mathbf{V}\left(t\right)=\sum_{k}\mathbf{V}_{k}e^{ik\Omega_dt}$. Note that we will assume that the driving strength coefficient is sufficiently small such that we can ignore terms of the order of $\mathcal{O}\left(\mathbf{V}_{j}^{3}\right)$ or higher.
Furthermore, since the Fourier coefficients $\mathbf{A}_{k}\left(\omega\right)$
are related to the Green function, one can interpret them as describing
the fundamental processes responsible for the phonon and hence heat
transport. These coefficients tell us that the driving is responsible for
a sudden change of the propagation frequency $\omega$ of the phonon
by an amount of $k\Omega_d$. Finally, the solution of the equations of motion
in this case would read as
\begin{equation}
X\left(t\right)=\sum_{k}\int_{-\infty}^{\infty}e^{-i\left(\omega-k\Omega_d\right)t}\mathbf{A}_{k}\left(\omega\right)\frac{\mathbf{B}\left(\omega\right)}{m}.
\end{equation}

\paragraph{Uncertainty relation.}

Finally, we comment that we check that the uncertainty relation holds for both cases that we consider, static and driven. This is simply expressed by the condition
\begin{equation}
\nu_{-}\geq\frac{1}{2}\label{eq:uncertainty principle},
\end{equation}
where $\nu_{-}$ is the minimum standard eigenvalue of $\widetilde{\mathbf{C}}\left(0\right)$
($\hbar$ is assumed to be equal to 1 in this case). In the above expression, $\widetilde{\mathbf{C}}\left(0\right)=i\mathbf{W}\cdot\mathbf{C}\left(0\right)$
with $\mathbf{W}$ the symplectic matrix and $\mathbf{C}\left(0\right)$ the covariance matrix, that both are defined in the appendix. Note that the covariance matrix can be expressed in terms of the Green's function and the spectral density, for both static and driven cases as is shown in the appendix.

\section{Heat current control between the BECs}
\label{sec:mes}

Here we present the thermodynamics quantities of interest, in order to evaluate the performance of our system as a heat current control platform
and as a thermal diode. These quantities cannot be expressed in terms of analytically known functions, due to the non-ohmic spectral density that describes the baths. As such, the results in next section are obtained by numerically evaluating the integrals.

\subsection{Static case}
We begin with the static scenario with $H_{\rm{Drive}}\left(t\right)=0$,
and we study the behaviour of heat current and the current-current correlations.
\paragraph{The heat current.}
When studying
a heat engine, a key quantity  is the average current $J_{\rm{L}}$ going from the left
reservoir (assuming it to be the hot reservoir) to the left oscillator
\cite{2008Dhar-1,2016Lepri} (by conservation of current $J_{\rm{L}}=-J_{\rm{R}}$),
or equivalently the average rate at which the left bath does work
on the left particle (power). In general, there are two ways to define
the heat current. The first  one is derived from
considerations of energy conservation on the system,  
\begin{equation}
J_{\rm{L}}=\frac{d\left\langle H_{\rm{S}}\right\rangle }{dt}-\left\langle \frac{\partial}{\partial t}H_{\rm{S}}\right\rangle =\left\langle \frac{i}{\hbar}\left[\widetilde{H}_{Int.,L},H'_{\rm{L}}\right]\right\rangle _{\rho},
\end{equation}
where $H'_{\rm{L}}=\frac{p_{\rm{L}}^{2}}{2m}+V_{\rm{L}}\left(x_{\rm{L}}\right)+\kappa x_{\rm{L}}x_{\rm{R}}$
and the average is over the total density state $\rho$. The second definition is expressed 
in terms of the rate of decrease of the bath energy~\cite{2016Kato}
\begin{eqnarray}
\hspace{-0cm}\widehat{J}_{\rm{L}}&=-\frac{d\left\langle H_{B,L}\right\rangle _{\rho}}{dt}=\left\langle \frac{i}{\hbar}\left[H_{B,L},H_{\rm{Tot}}\right]\right\rangle _{\rho}\label{eq:current defn bath energy}\\
&=\left\langle\frac{i}{\hbar}\left[\widetilde{H}_{Int.,L},H'_{\rm{L}}\right]\right\rangle_{\rho}+\frac{d\left\langle \widetilde{H}_{Int.,L}\right\rangle _{\rho}}{dt}+\left\langle \left[\widetilde{H}_{Int.,L},\widetilde{H}_{Int.,R}\right]\right\rangle _{\rho}.\nonumber
\end{eqnarray}
The second term vanishes under steady state conditions\textcolor{red}{{}
}and weak system-bath coupling, where weak is understood in the sense of assumption (\ref{eq:static bath assumption}) (and not of Markovian dynamics for the impurity),
which is the case in our study. We are further considering that the correlations among the system bath interactions is negligible, i.e. $\left\langle \left[\widetilde{H}_{Int.,L},\widetilde{H}_{Int.,R}\right]\right\rangle _{\rho_{\rm{B}}}=0$, since we assume that each system interacts only with its own reservoir. Under these criteria the two definitions of heat current are equivalent.
In other words, in our model, the rate at which the bath looses/gains energy is equal to the energy that the system gains/looses.
The more general scenario of strong coupling and hence non-separability
of the system-bath was considered for a spin-boson model
in~\cite{2015Gelbwaser}, while the case of interactions among the
baths was studied in~\cite{2012Li}. 
%
Therefore the heat current considered here is
\begin{equation}
J_{\rm{L}}=\left\langle \frac{i}{\hbar}\left[\widetilde{H}_{Int.,L},H'_{\rm{L}}\right]\right\rangle _{\rho_{\rm{B}}}=\left\langle \eta_{\rm{L}}\left(t\right)\frac{p_{\rm{L}}\left(t\right)}{m}\right\rangle _{\rho_{\rm{B}}},\label{eq:static_current_op}
\end{equation}
where the second equality is valid in steady state with $\eta_{\rm{L}}\left(t\right)=\int_{0}^{t}\lambda_{\rm{L}}\left(t-s\right)x_{\rm{L}}\left(s\right)\rm{d}s+\frac{1}{m}B_{\rm{L}}\left(t\right)$,
and it is obtained after solving the bath particles equations of motion.
Note that in (\ref{eq:static_current_op}) the current is 
a scalar, which
is the average current, and it is averaged under steady state condition
i.e. at the long time limit which is independent of the initial state
of the system. In~\cite{2008Dhar-1,2016Lepri,2003Dhar}, by using
a direct solution of the equations of motion for a non-interacting
system of bath particles, the average current is proven to be equal
to
\begin{equation}
\left\langle J_{\rm L}\right\rangle _{\rho_{\rm B}}=\frac{1}{4\pi}\int_{-\infty}^{\infty}{\rm d}\omega\mathcal{T}\left(\omega\right)\hbar\omega\left[f\left(\omega,T_{\rm{L}}\right)-f\left(\omega,T_{\rm{R}}\right)\right]=-\left\langle J_{\rm{R}}\right\rangle \eqqcolon\left\langle J\right\rangle, \label{eq:Landauer formula-1}
\end{equation}
with $f\left(\omega,T_{j}\right)=[{e^{\frac{\hbar\omega}{k_{\rm{B}}T_{j}}}-1}]^{-1}$, the phonon occupation number for the respective thermal reservoir
and
\begin{eqnarray}
&\mathcal{T}\left(\omega\right)=4Tr\left[\mathbf{G}\left(\omega\right)Im\left[\mathbf{L}_{\rm{L}}\left(\omega\right)\right]\mathbf{G}^{\dagger}\left(\omega\right)Im\left[\mathbf{L}_{\rm{R}}\left(\omega\right)\right]\right],\\
&\mathbf{L}_{\rm{L}}\left(\omega\right)=\left(\begin{array}{cc}
\lambda_{\rm{L}}\left(\omega\right) & 0\\
0 & 0
\end{array}\right),\:\mathbf{L}_{\rm{R}}\left(\omega\right)=\left(\begin{array}{cc}
0 & 0\\
0 & \lambda_{\rm{R}}\left(\omega\right)
\end{array}\right),
\end{eqnarray}
being the transmission coefficient for phonons at frequency $\omega$.
This is called the Landauer formula. 
Note that $\mathcal{T}\left(\omega\right)$ can be related to the
transmittance of plane waves of frequency $\omega$ across the system
as in~\cite{2012Das}. The transmission coefficient, interestingly,
depends on all the parameters of the system. That is, it depends on the parameters of the
baths (apart from temperature) as well as the parameters of the coupling
between the system and the baths. 

Equation~(\ref{eq:Landauer formula-1})
describes the situation where the central region is small in comparison
with the coherent length of the waves, which is the assumption we
also abide to, so that it is treated as purely elastic scattering
without energy loss. The dissipation resides solely in the heat baths.
This implies ballistic thermal transport, which corresponds to direct
point-to-point propagation of energy, contrary to the transport in
bulk and disordered structures which is referred to as diffusive transport.
Indeed, within the framework of modeling thermal baths by means of
quantum Langevin equations, ballistic transport has been observed
for chains of quantum harmonic oscillators~\cite{1990Zurcher}, and
hence this is also our case. Furthermore, note that the Landauer formula
can also capture phonon tunneling, i.e. the case when phonons off-resonance
with the systems vibrations cross the ``junction'', showing features
of quantum tunneling~\cite{2003Segal}. Finally, it is worth commenting
on an implicit assumption we made. We assumed here that a unique steady
state was reached, or equivalently that there are no bound states
in our system, i.e., that no modes outside the bath spectrum are generated
for the combined model of system and baths. The problem is that these
modes are localized near the system and any initial excitation of
the mode is unable to decay~\cite{2006Dhar}. 

It is interesting to comment on two limits of (\ref{eq:Landauer formula-1}),
namely the linear limit $\Delta T:=T_{\rm{L}}-T_{\rm{R}}\ll T$ where $T=\frac{T_{\rm{L}}+T_{\rm{R}}}{2}$
and the classical limit, $\frac{\hbar\omega}{k_{\rm{B}}T}\rightarrow0$ where $\omega$ refers to the bath frequencies. 
In the first limit, the current reduces to
\begin{equation}
\left\langle J\right\rangle =\frac{\Delta T}{2\pi}\int_{0}^{\infty}{\rm d}\omega\mathcal{T}\left(\omega\right)\hbar\omega\frac{\partial f\left(\omega,T\right)}{\partial T},\label{eq:Landauer linear limit}
\end{equation}
such that once the two baths are at the same temperature there is
no current flow. In the second limit, it becomes
\begin{equation}
\left\langle J\right\rangle =\frac{k_{\rm{B}}\Delta T}{2\pi}\int_{0}^{\infty}\rm{d}\omega\mathcal{T}\left(\omega\right),\label{eq:Landauer classical}
\end{equation}
where the current is independent of the
temperature of the baths $T_{\rm{L}},T_{\rm{R}}$ but it only depends on their
difference $\Delta T$.

We conclude with one last comment regarding our system and the role that entanglement  could play on the amount of heat current transported from one BEC to the other. From \cite{2019Riera-Campeny}, it is known that the static current can also be expressed in terms of the off-diagonal elements of the covariance matrix. This, might lead one to consider that entanglement, the presence of which is understood to be related to these off-diagonal elements might play a role in the amount of heat current transported. However, it was shown in \cite{2007Liu,   2010Ruggero} that in a system of two harmonic oscillators coupled to distinct baths, as is our case, there is no long-time entanglement present. 

\paragraph{Current-Current correlations.}

Next, we focus on the current-current correlations which is an easily accessible quantity from an experimental point of view. This is because these correlations are related to noise, which can be experimentally measured. They contain valuable information on the nature of the fundamental processes responsible for the heat transport. Furthermore, from the current-current correlation many other quantities can be obtained, such as the thermal conductance \cite{2008Dhar-1,2016Lepri} and the local effective temperature of driven systems \cite{2012Caso}.  


The current-current
time correlations $JJ_{\rm{LL}}\left(t,t'\right)$, which is sometimes referred to as current fluctuations
in time or current noise, is defined as the symmetrized correlation function of the current, that is
\begin{equation}
JJ_{\alpha\beta}\left(t,t'\right)=\frac{1}{2}\left\langle \left[J_{\alpha}\left(t\right)-\left\langle J_{\alpha}\left(t\right)\right\rangle _{\rho_{\rm{B}}},\:J_{\beta}\left(t'\right)-\left\langle J_{\beta}\left(t'\right)\right\rangle _{\rho_{\rm{B}}}\right]\right\rangle _{\rho_{\rm{B}}}.
\end{equation}
We are interested
in the steady state correlations, for which an expression for this
can be obtained using the non-equilibrium Green's functions as was
mentioned above. Hence the correlation function of interest is a function
only of the time difference, $JJ\left(t,t'\right)=JJ\left(t-t'\right)$. Therefore, 
the noise strength is characterized by the zero frequency component
$JJ_{\alpha\beta}=\int_{-\infty}^{\infty}JJ_{\alpha\beta}\left(t\right)\rm{d}t$, 
which obeys $JJ_{\alpha\beta}\geq0$ according to Wiener-Khinchine
theorem. It is current conserving, i.e. the sum of currents entering
the system from all reservoirs is equal to zero at each instant of
time, and gauge invariant and hence physically meaningful~\cite{2000Blanter,2005Kohler}.
Current conservation implies $JJ_{\rm{LL}}=JJ_{\rm{RR}}$.

One way to obtain
such an expression is by first deriving the cumulant generating function
$\chi\left(\mu\right)$, employing the non-equilibrium Green's functions
technique within the Keldysh formalism, and noting that
$JJ_{\alpha\beta}:=\frac{\partial^{2}\chi\left(\mu\right)}{\partial\mu^{2}}\mid_{\mu=0}$ (see e.g., ~\cite{2008Dhar-1,2016Lepri,2007Saito}).
In this case, one can show that the current fluctuations read as~\cite{2000Blanter,2005Kohler} 
\begin{eqnarray}
\hspace{-2.4cm}\left\langle JJ_{\rm{LL}}\right\rangle _{\rho_{\rm{B}}}=\int_{0}^{\infty}\rm{d}\omega\frac{\hbar^{2}\omega^{2}}{2\pi}\Bigg\{ \mathcal{T}\left(\omega\right)\left[f\left(\omega,T_{\rm{R}}\right)\left(1+f\left(\omega,T_{\rm{R}}\right)\right)+f\left(\omega,T_{\rm{L}}\right)\left(1+f\left(\omega,T_{\rm{L}}\right)\right)\right]\nonumber\\
-\mathcal{T}\left(\omega\right)\left(1-\mathcal{T}\left(\omega\right)\right)\left[f\left(\omega,T_{\rm{L}}\right)-f\left(\omega,T_{\rm{R}}\right)\right]^{2}\Bigg\}. 
\label{eq:Current Current corr-1}
\end{eqnarray}
The first two terms of this expression correspond to the equilibrium
noise, while the third corresponds to the non-equilibrium noise, also referred to as shot noise. At high energies, the latter is negligible. Note
that the shot noise is negative and hence contributes to diminish the
noise power in comparison with having the equilibrium noise alone.
The expression above Eq. (\ref{eq:Current Current corr-1}) is true
only under the assumption of independence of the two baths.  

Finally let us address the linear and the classical limits of the correlations.
In the first limit, the current-current correlations read as
\begin{equation}
\left\langle JJ_{\rm{LL}}\right\rangle _{\rho_{\rm{B}}}=\int_{0}^{\infty}\rm{d}\omega\frac{\hbar^{2}\omega^{2}}{2\pi}\left\{ \left[\mathcal{T}^{2}\left(\omega\right)\left(\Delta T\frac{\partial f\left(\omega,T\right)}{\partial T}\right)^{2}+\frac{2T^{2}\mathcal{T}\left(\omega\right)}{\omega}\frac{\partial f\left(\omega,T\right)}{\partial T}\right]\right\} .\label{eq:Current current linear}
\end{equation}
In the classical limit it becomes
\begin{equation}
\left\langle JJ_{\rm{LL}}\right\rangle _{\rho_{\rm{B}}}=\frac{k_{\rm{B}}^{2}}{2\pi}\int_{0}^{\infty}\rm{d}\omega Tr\left[\mathcal{T}^{2}\left(\omega\right)\left(\Delta T\right)^{2}+2\mathcal{T}\left(\omega\right)T_{\rm{L}}T_{\rm{R}}\right].\label{eq:current-current classical}
\end{equation}
Note that (\ref{eq:Current current linear}), contrary to the expression
for the current (\ref{eq:Landauer linear limit}) at the same limit,
does not vanish when $\Delta T\rightarrow0$, which results in nonzero
fluctuations of the current even in the scenario that no average current flows in the system. 

\subsection{The dynamic case}

For the driven case, the steady-state averaged heat current is given by~\cite{2019Riera-Campeny} 
\begin{align}
J_{j}^{\left(D\right)}&=-\int_{\mathbb{R}}\rm{d}\omega\widehat{T}_{j}\left(\omega\right)\left(f_{j}\left(\omega\right)+\frac{1}{2}\right)
\nonumber\\
&~~~+\sum_{q\neq j}\int_{\mathbb{R}}\rm{d}\omega\left[\widetilde{T}_{qj}\left(\omega\right)\left(f_{j}\left(\omega\right)+\frac{1}{2}\right)-\widetilde{T}_{jq}\left(\omega\right)\left(f_{q}\left(\omega\right)+\frac{1}{2}\right)\right],
\end{align}
where the new transmission coefficient reads as
\begin{equation}
\widetilde{T}_{jq}\left(\omega\right)=\sum_{k}\hbar\left(\omega-k\Omega_d\right)tr\left[Im\left[\mathbf{L}_{j}\left(\omega-k\Omega_d\right)\right]\mathbf{A}_{k}\left(\omega\right)Im\left[\mathbf{L}_{q}\left(\omega\right)\right]\mathbf{A}_{k}^{\dagger}\left(\omega\right)\right],
\end{equation}
and
\begin{equation}
\widehat{T}_{j}\left(\omega\right)=\sum_{\beta}\widetilde{T}_{jq}\left(\omega\right).
\end{equation}
This expression was also obtained in~\cite{2005Kohler,2017Freitas,2003Camalet,2004Camalet},
while in~\cite{2003Lehman} a similar expression was obtained for
transport through quantum dots. Unlike the static case, these transmission coefficients are not symmetric, i.e.,
$\widetilde{T}_{jq}\left(\omega\right)\neq\widetilde{T}_{qj}\left(\omega\right)$.
Crucially, this symmetry breaking, attributed to the driving that is now
expressed in the form of the transmission coefficients, is responsible for the appearance of {\it heat rectification}
as addressed in~\cite{2019Riera-Campeny}. 
To observe and quantify rectification, it is useful to evaluate the rectification
coefficient 
\begin{equation}
R:=\frac{\left|J_{j}^{\left(D\right)}+J_{j,\rm{r}}^{\left(D\right)}\right|}{\max\left(\left|J_{j}^{\left(D\right)}\right|,\left|J_{j,\rm{r}}^{\left(D\right)}\right|\right)},\label{eq:rectification coeff}
\end{equation}
where $J_{j,\rm{r}}^{\left(D\right)}$ is the value of the current,
once the temperature gradient is reversed, i.e. when the two baths'
temperatures are interchanged. Notice that this coefficient takes
values between 0 and 2, namely, $R=0$
when $J_{j}^{\left(D\right)}=-J_{j,\rm{r}}^{\left(D\right)}$, with the current being symmetric under reversing the temperature gradient. The upper bound is achieved when the current remains unaffected
by reversing the temperature gradient. When either of the two
currents is blocked, the coefficient is equal to one.

\section{Main Results}
\label{sec:results}


Before presenting our main results, let us summarize the major assumptions we made for our system, and the restrictions
that these impose on the parameter regimes that we can consider:
\begin{enumerate}
\item Linearization of the impurity-bath coupling, which is achieved by
assuming that the impurity is in the middle of its corresponding trap (see Eq. (\ref{eq:Linearity assumption})).
This in practice imposes a restriction on the maximum temperature
we can consider~\cite{2018Lampoa} 
\begin{equation}
T_{j}\ll T_{j}^{\rm{max}}=\frac{m\Omega_{j}^{2}R_{j}^{2}}{k_{\rm{B}}}.
\end{equation}
Note that in the scenario when the impurity is driven, $\Omega_{j}^{2}$
is replaced by $\min_{t}\left(\Omega_{j}^{2}+\sum_{k}V_{k,jj}e^{ik\Omega_dt}\right)$.
\item BEC independence condition $R_{\rm{L}}+R_{\rm{R}}<d$
\item Positivity condition, Eq. (\ref{eq:positivity condition}). 
\end{enumerate}
In our analysis below we consider the BECs of Rubidium (Rb) atoms. The impurities are Dysprosium (Dy) atoms, which are the atoms with the largest magnetic moment known at present, $\mu=10\mu_{\rm{B}}$, where $\mu_{\rm{B}}$ is the
Bohr magneton.

\subsection{Static system}
\begin{table}
\begin{center}
\begin{tabular}{ |p{4.5cm}|p{4.5cm}||p{3.5cm}|  }
	\hline
	Left BEC and impurity &Right BEC and impurity&Other parameters\\
	\hline
	$T_{\rm{L}}=75nK$ &$T_{\rm{R}}=7.5nK$&  $d=9.5a_{\rm{ho}}$\\
	$N_{\rm{L}}=7.5\times 10^{4}$ & $N_{\rm{R}}=7\times 10^{4}$ &$a_{\rm{ho}}=0.7\mu m$\\
	$\eta_{\rm{L}}=0.5$ & $\eta_{\rm{R}}=0.5$ &$\Omega=200\pi Hz$\\
	$g_{\rm{B}}^{\left(L\right)}=2.5\times 10^{-40}J\cdot m$ & $g_{\rm{B}}^{\left(R\right)}=2\times10^{-40}J\cdot m$ & \\
	$\Omega_{\rm{L}}=\Omega$ & $\Omega_{\rm{R}}=\Omega$&\\
	$\widehat{\Omega}_{B,L} = 3\Omega$ & $\widehat{\Omega}_{B,R}=3\Omega$ &\\
	\hline
\end{tabular}	
\caption{List of default parameters---unless otherwise mentioned---used in the static case, corresponding to Figures.~\ref{fig:fig2} and ~\ref{fig:fig3}.\label{table:tab1}}
\end{center}
\end{table}
%
%
%
\subsubsection{Heat current.}

\begin{figure}
\centering{}%
\includegraphics[width=0.99\textwidth]{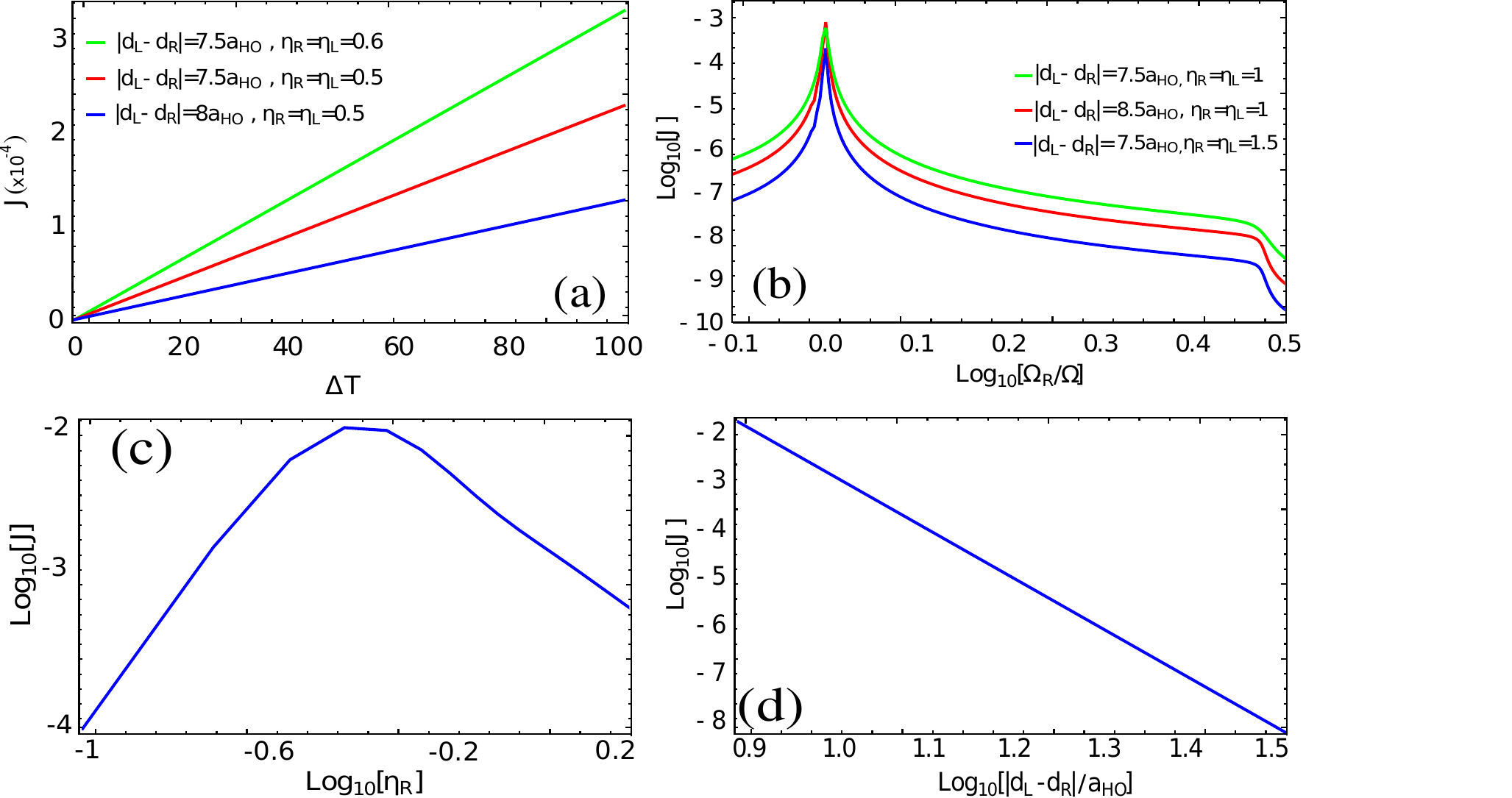}
\caption{{\bf(a)} Heat current $J$ against the temperature difference of the two baths $\Delta T$, with fixed $T_{\rm R}$. As expected current
increases linearly with $\Delta T$.
Furthermore, the current decreases with increasing distance, and increases with increasing the coupling strength of the
impurities to the bath. \textbf{(b) } 
Current as a function
of the trapping frequency $\Omega_{\rm{R}}$ of the right impurity.
We observe resonance at $\Omega_{\rm{R}} = \Omega_{\rm{L}}$. Furthermore, for $\Omega_{\rm R}$
larger than the trapping frequency of the bath---which is also the cutoff for the
spectral density---the current vanishes quickly. Moreover, current decreases with distance
as before. In this regime, current decreases as the
impurities couple stronger to their respective baths (note that the coupling strengths are different from panel {\bf (a)}). \textbf{(c)} Current as a function of the coupling strength of the right impurity. Current reaches a maximum in the range studied.  \textbf{(d)} Current as a function of the distance between the impurities $|d_1-d_2|$. Current decreases linearly with increasing distance in the regime we were allowed to study, that is under the restriction that is imposed on the lower distance in order to maintain the two BECs spatially independent. Current is found to scale as $\kappa^{2}$. See Table~\ref{table:tab1} for the parameters that we use here. \label{fig:fig2}}
\end{figure}

In Fig.~\ref{fig:fig2} (a) we plot the heat current with the temperature difference $\Delta T=T_{\rm{L}}-T_{\rm{R}}$, where we keep fixed the temperature of the left reservoir $T_{\rm{L}}$. 
As expected, the heat current increases with increasing temperature difference, while it is zero when there is no temperature gradient. We see that the current depends linearly on the temperature gradient. This is in accordance to the linear limit for the heat current in Eq.~(\ref{eq:Landauer linear limit}). 
 In addition, we studied heat current in the scenario where the temperature difference between
the two baths was fixed to some value, in particular $\Delta T=10nK$,
with $T_{\rm{L}}=T$ and $T_{\rm{R}}=T+\Delta T$, and we considered the simultaneous
variation of the temperatures of both baths, in the temperature regime
$T=10{\rm nK}-100 {\rm nK}$. In this case we saw 
that the heat current remained
constant as a function of $T$ and hence we conclude that the regime
in which we could study the system was that of the classical limit
Eq. (\ref{eq:Landauer classical}). A figure for this case is omitted since the current was just constant as a function of $T$. From Fig.~\ref{fig:fig2} (a) we also observe
that increasing the distance of the two impurities results in decreasing the heat current
flow (red {\it vs} blue curves), while increasing the impurity-BEC couplings results in an increase of the current (red {\it vs} green curves). 

Figure~\ref{fig:fig2} (b) depicts the heat current against
the trapping frequencies of the BECs. In particular we fix
one of the trapping frequencies and vary the other. 
Firstly, we observe that the heat current is reaches a maximum when the trapping frequencies of the two impurities match, i.e., $\Omega_{\rm R} = \Omega_{\rm L}$. This is understood as follows.
The current density, which we define as $J_{den.}:=\frac{1}{4\pi}\mathcal{T}\left(\omega\right)\hbar\omega\left[f\left(\omega,T_{\rm{L}}\right)-f\left(\omega,T_{\rm{R}}\right)\right]$
is maximized whenever the denominator of the transmission function
$\mathcal{T}\left(\omega\right)$, given by $\left(-\omega^{2}\mathbf{I}+\mathbf{K}-\mathbf{L}\left(\omega\right)\right)^{-1}\left(\left(-\omega^{2}\mathbf{I}+\mathbf{K}-\mathbf{L}\left(-\omega\right)\right)^{T}\right)^{-1}$
is minimized. In the regime we are looking, $\frac{\tau_{j}}{\widehat{\Omega}_{j}^{3}}$
come out to be of the order of $10^{-4}$, while the values of $\kappa$
that are allowed, are of the order of $10^{-5}$. These are much smaller
than the trapping frequencies $\Omega_{\rm{L}}^{2}$, $\Omega_{\rm{R}}^{2}$,
such that the denominator is minimized whenever $\left(\omega^2-\Omega_L^2\right)^2+\left(\omega^2-\Omega_R^2\right)^2$ is minimized. This happens when $\omega=\Omega_L=\Omega_R$. 
Secondly---for the specific parameters that we choose---contrary to Fig.~\ref{fig:fig2} (a), 
increasing the impurity-BEC coupling strength results in reducing the
current. 

We study the dependence on the impurity-BEC coupling strength in Fig.~\ref{fig:fig2} (c), where we see current reaches a maximum at some optimal coupling. Keeping the coupling constant of the left impurity fixed and varying that of the right impurity, we find that if the impurity is weakly coupled to the BECs, then current can not be carried from one BEC to the other through the vibrations of these  impurities. 
 If on the other hand, this is coupled too strongly, the effect of the noise induced by the baths (BECs) reduces the current that can be transmitted 
This is in agreement with the findings in Fig.~\ref{fig:fig2} (a) and (b). 

In Fig.~\ref{fig:fig2} (d) we see how the heat current varies as a function of the distance between the two impurities. As expected, we see that increasing the distance between the impurities reduces the heat current. In particular, we find $\left\langle J\right\rangle \propto\kappa^{2}$---in the parameter regime that we study.
It is possible that, at shorter distances, another resonance effect occurs between the value of the spring constant, which depends inversely on the distance between the impurities and the trapping frequencies of the impurities. Anyhow, we do not study the system  at such small distances because the approximation of independence of the BECs breaks down. 

Finally we remark here that one could also consider studying homogeneous
gases instead of harmonically trapped ones and the induced heat current
could be examined for this case, by studying the system in the spirit
of~\cite{2017Lampo}. Experimentally, homogeneous gases could be
created as in~\cite{2013Gaunt}. From our studies, we observe that
as expected one can still have current in this case, but since the
spectral density is different in this case, even though still superohmic,
there is a quantitative difference in the amount of current.  Nevertheless
we focused on the harmonically trapped case which is more conventionally
implemented experimentally. From~\cite{2018Lampoa} it is known that
the two spectral densitites result in a different degree of non-markovianity,
and it would be interesting in the future to study the effect of non-markovianity
on the heat current. This however exits the scope of this
paper. 

\subsubsection{Current-Current correlations.}

\begin{figure}
\includegraphics[width=0.99\textwidth]{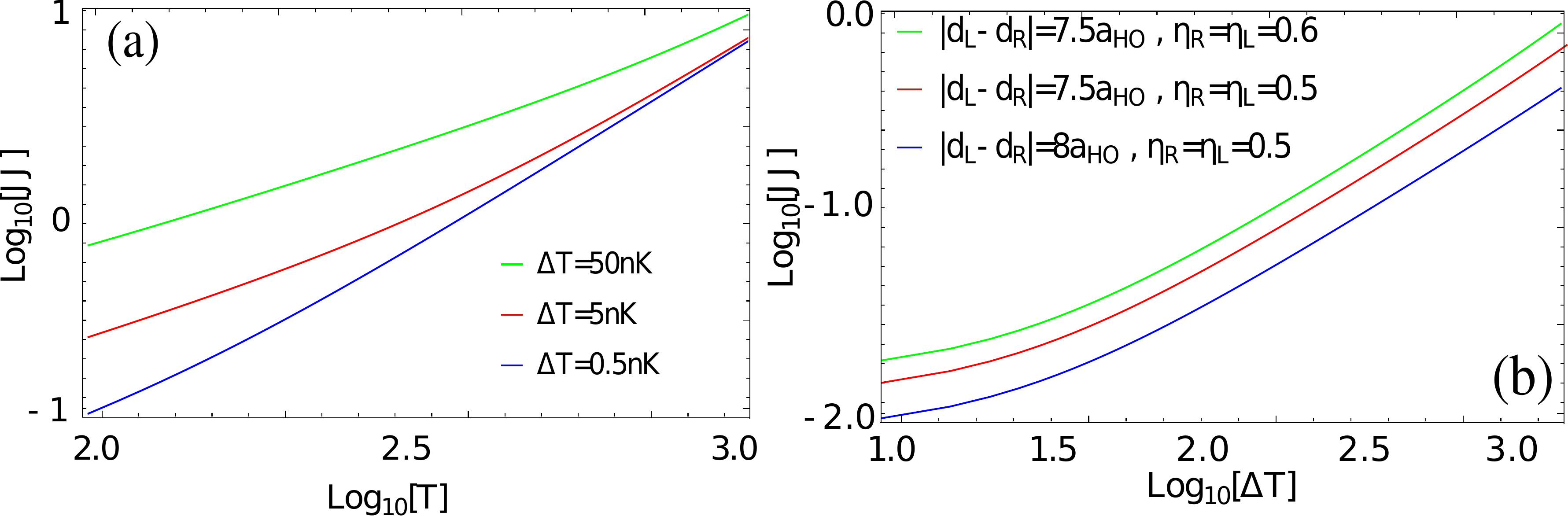}
\caption{
\textbf{(a) } Current-current
correlations against temperature $T$ at constant $\Delta T$. Here we observe that for small $\Delta T$ and at large temperatures
$T$ as expected from Eq. (\ref{eq:current-current classical}), current-current
correlations scale as $\propto T^{2}$. At sufficiently large $T$
this is expected to happen for large $\Delta T$ as well, however
we were restricted on the range of maximum temperatures we could consider. \textbf{(b)} Current-current correlations
as a function of the temperature difference of the two baths $\Delta T$.
For small values of $\Delta T$,
i.e., in the linear regime, the correlations have a nonzero value as
predicted in Eq. \eqref{eq:Current current linear}. This implies
that even as $\Delta T\rightarrow0$, where current also vanishes,
correlations still persist. For large enough values of $\Delta T$ the correlations increase linearly with $\Delta T$. See Table~\ref{table:tab1} for the parameters that we use here.}  \label{fig:fig3}
\end{figure}

In Fig.~\ref{fig:fig3} (a) the behavior of the current-current correlations is illustrated as a function of temperature $T$ while keeping the
temperature difference $\Delta T$ constant.  From the figure, we see that, for small
$\Delta T$, such that the first term of Eq. (\ref{eq:current-current classical})
prevailed, the current-current correlations are proportional to $T^2$. On the contrary when $\Delta T$ is
large, and for relatively small $T$, the current-current correlations
depended linearly on the temperature $T$.
Nevertheless the behavior
seems to be independent of $\Delta T$ as the temperature increases,
and appears to depend on the square of $T$ as expected in the classical
limit. 

In Fig.~\ref{fig:fig3} (b) we study the dependence of current-current correlations on the temperature difference $\Delta T$. At large temperature difference, i.e. beyond the linear limit considered in (\ref{eq:Current current linear}), the correlations depend linearly on $\Delta T$. We comment here that this is not directly evident from the LogLog plot, but we checked that this is indeed the case of a linear relation from the corresponding current-current correlation versus temperature difference plot before taking the logarithms. 
For small $\Delta T$, where (\ref{eq:Current Current corr-1})
is well approximated by Eq.~(\ref{eq:Current current linear}), indeed
we find the saturation on the correlations predicted by the second
term of Eq.~(\ref{eq:Current Current corr-1}). This implies that
even as $\Delta T\rightarrow0$---in which case current is zero---
the fluctuations are still present. One might expect quantum effects
to appear in this regime then, but we studied the entanglement in this system,
by means of the logarithmic negativity as in~\cite{2019Charalambous},
and no entanglement could be detected in this regime. 


\subsection{Driven case: Heat rectification}
\begin{table}
	\begin{center}
		\begin{tabular}{ |p{4.5cm}|p{4.5cm}||p{3.5cm}|  }
			\hline
			Left BEC and impurity &Right BEC and impurity&Other parameters\\
			\hline
			$T_{\rm{L}}=15nK$ &$T_{\rm{R}}=1nK$&  $d=35a_{\rm{ho}}$\\
			$N_{\rm{L}}=7.5\times 10^{4}$ & $N_{\rm{R}}=7\times 10^{4}$ &$a_{\rm{ho}}=0.7\mu m$\\
			$\eta_{\rm{L}}=0.5$ & $\eta_{\rm{R}}=0.5$ &$\Omega=200\pi Hz$\\
			$g_{\rm{B}}^{\left(L\right)}=5\times 10^{-39}J\cdot m$ & $g_{\rm{B}}^{\left(R\right)}=4.5\times10^{-39}J\cdot m$ &$v=0.1\Omega$ \\
			$\Omega_{\rm{L}}=2 \Omega$ & $\Omega_{\rm{R}}=2 \Omega$&\\
			$\Omega_{B,L} = 2.5 \Omega$ & $\Omega_{B,R}= 2.5 \Omega$ &\\
			\hline
		\end{tabular}	
		\caption{List of default parameters---unless otherwise mentioned---used in the dynamic case, corresponding to Fig.~\ref{fig:fig4}}
		\label{table:tab2}
	\end{center}
\end{table}

Heat
rectification is quantified by the heat rectification, Eq.~(\ref{eq:rectification coeff}). We use the particular
form of driving 
\begin{equation}
\mathbf{V}\left(t\right)=2v\cos(\omega_dt)\left(\begin{array}{cc}
1 & 0\\
0 & 0
\end{array}\right),
\end{equation}
i.e. we drive the frequency of the first oscillator. 
The parameters that we use are given in Table~\ref{table:tab2}.
The temperatures we choose are upper bounded by $\{T_L, T_R\} < 10^3 nK$.

\begin{figure}
\centering{}
\includegraphics[width=0.7\textwidth]{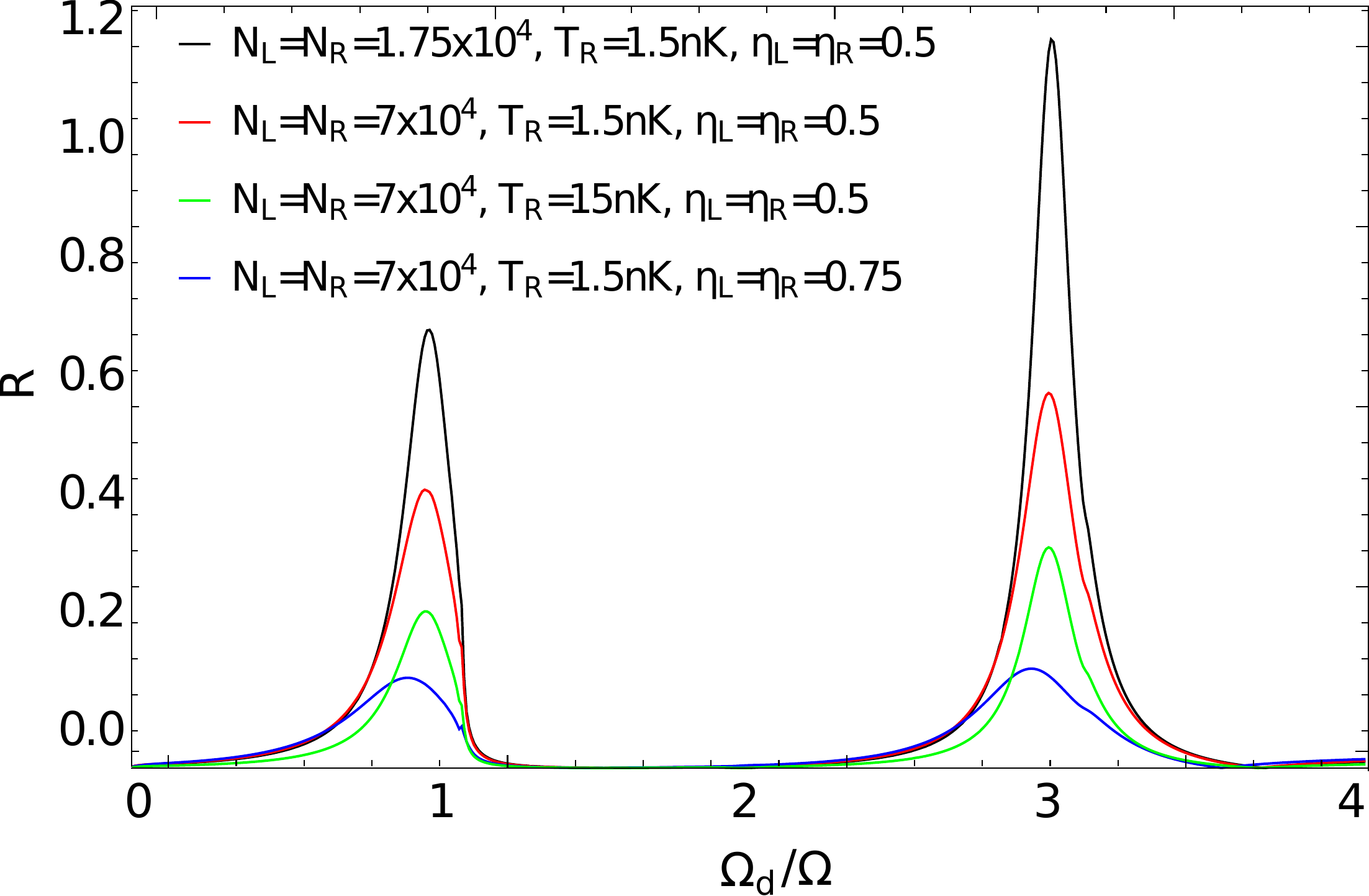}
\caption{Rectification coefficient against the driving
frequency. As predicted
analytically, we observe nonzero rectification at the vicinity of frequencies $\Omega_d=\{\Omega,3\Omega\}$
corresponding to $\left|\nu_{j}\pm\nu_{i}\right|$. Furthermore, we see that rectification decreases with
increasing coupling to the baths in this regime of parameters. See Table~\ref{table:tab2} for the parameters that we use here. 
} \label{fig:fig4} 
\end{figure}

In Fig.~\ref{fig:fig4} we depict the heat rectification coefficient $R$ as a function of $\Omega_d$, the driving frequency. We find that it shows to maxima at for two values of the driving frequency
$\Omega_d\in\{\Omega, 3\Omega\}$. Note that, as shown in~\cite{2019Riera-Campeny}, heat rectification should be maximum at the following frequencies 
\begin{equation}
\Omega_d=\left|\nu_{j}\pm\nu_{i}\right|.\label{eq:normal mode Rect}
\end{equation}
Here $\{i,j\}\in\left\{\rm{L,R}\right\}$ with $i\neq j$ and $\nu_i$'s are the normal modes of the coupled impurities
\begin{equation}
\nu_{\rm{L,R}}^{2}=\Omega_{\rm{L}}^{2}+\kappa+\frac{\Delta}{2}\mp\left(\kappa^{2}+\frac{\Delta^{2}}{4}\right)^{1/2},
\end{equation}
with $\Delta=\Omega_{\rm{R}}^{2}-\Omega_{\rm{L}}^{2}$. 
In our case, Eq. (\ref{eq:normal mode Rect})
indeed suggests that rectification should be maximum at $\Omega_d\in\{\Omega, 3\Omega\}$ which explains the results  in Fig.~\ref{fig:fig4}. Note that we also studied dependence of
the rectification coefficient on the other parameters of the system,
apart from the driving frequency. In particular, we find that in the
regime of parameters we study, maximum rectification decreases when the impurities
couple more strongly to their respective baths. On the contrary, decreasing
the number of atoms significantly increases $R$,
in fact reaching $R>1$. What is more, rectification
could also be optimized with respect to the detuning between the trapping
frequencies of the impurities as in~\cite{2019Riera-Campeny}. 
%

\section{Conclusions}
\label{sec:conc}

In this work, we studied in detail the heat transport control and heat current rectification
among two Bose Einstein Condensates (BEC).
To this aim, we took an open quantum system approach, and focused on experimentally realistic conditions and
parameter regimes. In particular we considered a system composed of
two harmonically trapped interacting impurities immersed in two independent
harmonically trapped 1D BECs kept at different temperatures. The impurities
interact through a long range interaction, in particular
a dipole-dipole coupling---that under suitable conditions we were able
to treat as a spring-like interaction. We showed the dynamics
of these impurities can be described within the framework of quantum Brownian motion, where the excitation modes of the gas play the role of the bath. In this analogy, the spectral density of the bath is not postulated, but it is rather derived exactly from the Hamiltonian of
the BEC, which turns out to be superohmic. 
By solving the relevant generalized Langevin equations, we find the steady state covariance matrix of the impurities, which contains all the information describing our Gaussian system. 
In particular, we use such information to study the heat currents and current-current correlations and their dependence on the controllable parameters of the system. 
We find that, the heat current scales linearly with temperature difference among the two BECs. Furthermore, we observe that heat current is maximum when the trapping frequencies of the impurities are at resonance. Finally, we showed the
existence of an {\it optimal} coupling strength of the impurities on their respective baths.

What is more, by periodically driving one of the impurities, we can conduct heat asymmetrically, i.e., we achieve heat rectification---which is in full agreement with the recent proposal of~\cite{2019Riera-Campeny}. 
In particular, we see that one can achieve heat rectification at the driving frequencies predicted in~\cite{2018Renklioglu}, even though our bath is superohmic. 

Motivated by recent developments
on the usage of BECs as platforms for quantum information processing,
as e.g. in~\cite{2019Charalambous}, our work offers an alternative
possibility to use this versatile setting for information transfer
and processing, within the context of phononics. The possibility of quantum advantages using many-body impurities in our platform remains an interesting open question (see ~\cite{2016Jaramillo} too). Another future direction is to study heat control in 2D and 3D BECs. In principle this gives rise to a different spectral density, which opens a new window for further manipulation of heat current. Finally, it is desirable to investigate scenarios where the system is nonlinear, which raises difficulties in solving the problem analytically, nonetheless, it offers the opportunity to rectify heat even without periodic derivation. Moreover, motivated by the results in \cite{2019aMehboudi,2019bMehboudi} where the squeezing in position of a  single impurity embedded in a BEC was used to measure the temperature of the BEC in the sub-nano-Kelvin regime, one may study if the present two-particle set-up can be used for applications in quantum thermometry.


\section*{Acknowledgements}
We thank fruitful discussions with Andreu Riera-Campeny. 
We acknowledge  the Spanish Ministry MINECO (National Plan 15 Grant: FISICATEAMO No. FIS2016-79508-P, SEVERO OCHOA No. SEV-2015-0522), the Ministry of Education of Spain (FPI Grant BES-2015-071803), European Social Fund, Fundaci\'o Cellex, Generalitat de Catalunya (AGAUR Grant No. 2017 SGR 1341 and CERCA/Program), ERC AdG OSYRIS and NOQIA, EU FETPRO QUIC, and the National Science Centre, Poland-Symfonia Grant No. 2016/20/W/ST4/00314. MM acknowledges support from the Spanish MINECO (QIBEQI FIS2016-80773-P and Severo Ochoa SEV-2015-0522), Fundaci\'o Privada Cellex, and the Generalitat de Catalunya (CERCA Program and SGR1381).

\appendix
\section{The uncertainty relation}
\label{sec:App}
Here, we present how the covariance matrix of our system is obtained, (both for the static and driven cases) which we used in order to verify that our system fulfils the uncertainty principle. 
This allows us to guarantee that the system under
study, for the parameters considered and according with the assumptions that we have imposed, is a physical
system. To do so, one needs to obtain the covariance matrix of the system,
\begin{equation}
\mathbf{C}\left(0\right)=\left(\begin{array}{cc}
C_{\mathbf{XX}}\left(0\right) & C_{\mathbf{XP}}\left(0\right)\\
C_{\mathbf{PX}}\left(0\right) & C_{\mathbf{PP}}\left(0\right)
\end{array}\right),\label{eq:Covariance matrix}
\end{equation}
where $\mathbf{P}=\left(p_{1}\left(t\right),p_{2}\left(t\right)\right)^{T}$
and 
\begin{equation}
C_{\mathbf{AB}}\left(t-t'\right)=\frac{1}{2}\left\langle \mathbf{A}\left(t\right)\mathbf{B}^{T}\left(t'\right)+\mathbf{B}\left(t\right)\mathbf{A}^{T}\left(t'\right)\right\rangle _{\rho_{\rm{B}}}.
\end{equation}
We emphasize that the state of the system-bath is assumed to be a product state
as in (\ref{eq:static bath assumption}). Hence the average is
taken over the thermal state of the bath, while the state of the system
is assumed to have reached its unique equilibrium state by considering
the long-time, steady state limit $t\rightarrow\infty$ which is equivalent
to consider that $\omega\ll\Lambda_{\rm{L}},\Lambda_{\rm{R}}$. The $4\times 4$ matrix
in (\ref{eq:Covariance matrix}) is constructed from the vector $\mathbf{Y}\left(t\right)=\left(x_{1}\left(t\right),x_{2}\left(t\right),p_{1}\left(t\right),p_{2}\left(t\right)\right)^{T}$
as the product $\frac{1}{2}\left\langle \left\{ \mathbf{Y}^{T}\left(t\right)\mathbf{Y}\left(t'\right),\left(\mathbf{Y}^{T}\left(t\right)\mathbf{Y}\left(t'\right)\right)^{T}\right\} \right\rangle _{\rho_{\rm{B}}}$. 

The uncertainty relation will then be expressed as a condition on
the symplectic transform of the covariance matrix $\widetilde{\mathbf{C}}\left(0\right)$,
where the latter is obtained as $\widetilde{\mathbf{C}}\left(0\right)=i\mathbf{W}\cdot\mathbf{C}\left(0\right)$
with $\mathbf{W}$ the symplectic matrix $\mathbf{W}=\left(\begin{array}{cccc}
0 & 0 & 1 & 0\\
0 & 0 & 0 & 1\\
-1 & 0 & 0 & 0\\
0 & -1 & 0 & 0
\end{array}\right)$. The uncertainty relation then simply reads as
\begin{equation}
\nu_{-}\geq\frac{1}{2}\label{eq:uncertainty principle},
\end{equation}
where $\nu_{-}$ is the minimum standard eigenvalue of $\widetilde{\mathbf{C}}\left(0\right)$
($\hbar$ is assumed to be equal to 1 in this case).

\paragraph{Static covariance matrix.}

The elements of the covariance matrix in the static case, in terms of the phonons
propagator functions, can be expressed as 
\begin{eqnarray}
\hspace{2.cm}C_{x_{j}x_{q}}\left(0\right)=\frac{\hbar}{2\pi}\int_{-\infty}^{\infty}\rm{d}\omega \, Z_{jq}\left(\omega\right),\\
\hspace{2.cm}C_{x_{j}x_{q}}\left(0\right)=\frac{\hbar}{2\pi}\int_{-\infty}^{\infty}\rm{d}\omega \, i m\omega Z_{jq}\left(\omega\right),\\
\hspace{2.cm}C_{x_{j}x_{q}}\left(0\right)=\frac{\hbar}{2\pi}\int_{-\infty}^{\infty}\rm{d}\omega \, m^{2}\omega^{2}Z_{jq}\left(\omega\right),
\end{eqnarray}
where
\begin{equation}
Z_{jq}\left(\omega\right)=\sum_{k,s=1}^{2}\left(\mathbf{G}\left(\omega\right)\right)_{js}\left(Im[\mathbf{L}\left(\omega\right)]\cdot\coth\left[\frac{\hbar\omega}{2k_{\rm{B}}\mathbf{T}}\right]\right)_{sk}\left(\mathbf{G}\left(-\omega\right)\right)_{kq},
\end{equation}
with $\mathbf{T}=\left(\begin{array}{cc}
T_{\rm{L}} & 0\\
0 & T_{\rm{R}}
\end{array}\right)$ being the temperatures of each bath. 


\paragraph{Driven case.}

In the driven case, since the solution of the equations of motion is periodic,
so are the elements of the covariance matrix as well. In particular, in terms of the
Fourier components of the expansion of the periodic phonon Green function,
the covariance matrix elements read as
\begin{eqnarray}
&C_{x_{j}x_{q}}\left(0\right)=\frac{\hbar}{2\pi}\int_{-\infty}^{\infty}\rm{d}\omega\sum_{k,l}\widetilde{Z}_{jq,kl}\left(\omega\right),\\
&C_{x_{j}x_{q}}\left(0\right)=\frac{\hbar}{2\pi}\int_{-\infty}^{\infty}\rm{d}\omega\sum_{k,l}im\left(\omega-l\Omega_d\right)\widetilde{Z}_{jq,kl}\left(\omega\right),\\
&C_{x_{j}x_{q}}\left(0\right)=\frac{\hbar}{2\pi}\int_{-\infty}^{\infty}\rm{d}\omega\sum_{k,l}m^{2}\left(\omega-k\Omega_d\right)\left(\omega-l\Omega_d\right)\widetilde{Z}_{jq,kl}\left(\omega\right),
\label{eq:Covariance driven}
\end{eqnarray}
where
\begin{equation}
\widetilde{Z}_{jq,kl}\left(\omega\right)=\sum_{n,s=1}^{2}\left(\mathbf{A}_{k}\left(\omega\right)\right)_{js}\left(Im[\mathbf{L}\left(\omega\right)]\cdot\coth\left[\frac{\hbar\omega}{2k_{\rm{B}}\mathbf{T}}\right]\right)_{sn}\left(\mathbf{A}_{\rm{L}}\left(-\omega\right)\right)_{nq}e^{i\left(k-l\right)\omega_dt}.
\end{equation}

\bibliographystyle{iopart-num}
\bibliography{herebiblio}

\end{document}